\documentclass{ametsocV6.1}

\usepackage{verbatim}
\usepackage{xspace}
\usepackage{csquotes}
\newcommand{\El}{El Ni\~{n}o\xspace}

\newcommand{\degree}{$^\circ$}

\title{How do the Pacific and Atlantic Oceans synergize to modulate Southeastern United States Precipitation Variability?}

\authors{Priyanshi Singhai,\aff{a}\correspondingauthor{Priyanshi Singhai, priyanshi.singhai-1@ou.edu} 
Kathy Pegion,\aff{a} 
Akintomide A. Akinsanola,\aff{b,c} 
Bohar Singh,\aff{d} 
Thierry N. Taguela,\aff{b}
}

\affiliation{\aff{a}{School of Meteorology, University of Oklahoma, Norman, OK, USA}\\
\aff{b}{Department of Earth and Environmental Sciences, University of Illinois Chicago, IL, USA}\\
\aff{c}{Environmental Science Division, Argonne National Laboratory, Lemont, IL, USA}\\
\aff{d}{NOAA Climate Prediction Center (CPC), College Park, Maryland, USA}
}

\abstract{This study explores the mechanisms behind anomalous positive and negative rainfall events in the southeastern United States (SEUS), emphasizing the interplay between upper-level large-scale atmospheric teleconnections and the lower-level North Atlantic Subtropical High (NASH). Through a novel conditional weather regime analysis of geopotential height at both lower and upper levels across the Pacific-North America-Atlantic region, we identify distinct clusters representing persistent and recurring circulation patterns originating from the Pacific and Atlantic Oceans. Our analysis of lower-level conditional weather regimes reveals two distinct phases of the NASH that influence rainfall patterns in the SEUS region. In one phase, the weakening and eastward shift of the NASH’s northern boundary reduces the central low-level jet, enhances cyclonic circulation, and increases rainfall in the SEUS. In the other phase, the excessive latent heating associated with enhanced SEUS rainfall triggers a wave train pattern that strengthens the intensity of NASH. Conversely, the opposite conditions apply during anomalous negative rainfall events. Additionally, the upper-level conditional weather regime indicates that large-scale dynamics of East Asian summer monsoons trigger the Rossby wave patterns, contributing considerably to the variability in SEUS rainfall from the upper levels. Therefore, our research highlights the crucial role of global atmospheric teleconnections at upper and lower levels in shaping SEUS precipitation patterns.
}

\begin{document}

\maketitle

%
%
%
\statement
This study examines the causes of unusual high and low rainfall in the southeastern United States, focusing on how large-scale weather patterns from the Pacific and Atlantic Oceans affect local precipitation. We identify key recurring weather patterns that contribute to these rainfall variations. A key finding is that the high-pressure system over the Atlantic Ocean, along with monsoonal rains in subtropical East Asia, influences local weather through large-scale circulation patterns. This underscores how weather systems from both oceans interact to influence rainfall in the region.

%
%
%

%







\section{Introduction}
During the boreal summer, the Southeastern United States (SEUS) experiences abundant precipitation with high subseasonal variability \citep{wei2019intraseasonal}. This anomalous precipitation, particularly extreme precipitation events, greatly affects the region's agriculture, water resources, energy, and economy \citep{seager2009drought,gotvald2010epic,ingram2013climate,eck2020influence}. Therefore, accurately predicting precipitation at a subseasonal timescale is critically important for the SEUS region. However, predicting precipitation in this region, particularly in summer, poses significant challenges. This is because summer precipitation is primarily driven by internal atmospheric variability \citep{seager2009drought} and influenced by various weather and synoptic-scale convective systems, which have low predictability during this season \citep{infanti2014southeastern}. In addition, large-scale climate drivers also modulate circulation patterns and impact moist convection over the SEUS region through various atmospheric and oceanic teleconnections. Therefore, understanding these linkages at subseasonal timescales is crucial for identifying the causes of anomalous precipitation in the SEUS region.

Several hypotheses have been proposed to explain the subseasonal-to-seasonal (S2S) variability of summertime precipitation in the SEUS region. However, most of these theories are based on an understanding of seasonal variability. These theories include but are not limited to internal atmospheric variability and large-scale atmospheric circulation patterns \citep{katz2003stochastic,knight2009contribution,seager2009drought,wei2019intraseasonal}, sea surface temperature (SST) anomalies and their associated teleconnections \citep{wang2001tropical,seager2003air,schubert2009us,kushnir2010mechanisms}, and hurricane activity \citep{elsner1993complexity}. Despite the significant importance of subseasonal variability and its impacts, little attention has been given to understanding the summer subseasonal variability of precipitation in the SEUS region.

Among all these factors, the North Atlantic Subtropical High (NASH) has been identified as the primary driver in controlling the S2S variability of summer SEUS precipitation \citep{davis1997north,li2011changes,li2012variation}. NASH is a semi-permanent high-pressure system located over the North Atlantic Ocean in the lower troposphere (850 hPa), which intensifies during the summer months with its center near Bermuda \citep{seager2003air,nigam2009summertime}. On a seasonal timescale, rainfall in the SEUS increases when the western ridge of the NASH is positioned southwest of its climatological location due to NASH's intensification. In contrast, a precipitation deficit occurs when the western ridge shifts northwest. Consequently, the position and intensity of NASH's western ridge play a critical role in transporting warm, moist air from the subtropical Atlantic Ocean and the Gulf of Mexico, influencing moist convection over the SEUS region \citep{li2012variation}. On a subseasonal timescale, the SEUS rainfall has a spectral peak at a 10–20-day timescale and is influenced by three-way dynamic interactions between a weakened NASH, a reduced central US low-level jet, and subsequent changes in moisture convergence and cyclonic circulation over SEUS \citep{wei2019intraseasonal}.

Another set of theories emphasizes the influence of SST anomalies in the Pacific Ocean on winter precipitation in the SEUS, with minimal attention given to summer precipitation patterns. It is well-known that the \El Southern Oscillation (ENSO) significantly affects winter precipitation in the SEUS through the Pacific-North American (PNA) teleconnection \citep{horel1981planetary,wallace1981teleconnections,barnston1987classification}. During winter, the westerly jet stream is stronger and positioned farther south than in summer, allowing it to be perturbed by the tropical/eastern Pacific heating associated with \El, thereby impacting rainfall in the SEUS region. In contrast, during summer, ENSO's impact on SEUS precipitation is generally weak and insignificant due to the jet stream's weaker state and northward shift \citep{li2013atmospheric,zhu2016new}. However, the East Asian monsoon system, extending to 40\degree N, interacts with the northward-shifted westerly jet stream over subtropical East Asia. The monsoonal heating over this region can trigger Rossby waves that travel along the westerly jet stream from East Asia to North America, forming the Asia-North America (ANA) teleconnection. This wave propagation influences summer rainfall patterns across the United States \citep{zhu2016new}. Furthermore, North Atlantic SST anomalies have been identified as having a greater impact on SEUS rainfall than ENSO on both annual \citep{wang2010intensification} and decadal timescales \citep{hu2011variations}. Additionally, \cite{curtis2008atlantic} pointed out that the Atlantic Multidecadal Oscillation (AMO) might influence SEUS summer precipitation through its effect on hurricane activity. 

All these factors may potentially influence S2S summer precipitation in the SEUS by interacting at various spatial and temporal scales. For example, the increased seasonal variability in SEUS precipitation in recent decades can be attributed to the combined effects of the intensification of the NASH and a positive Pacific Decadal Oscillation (PDO) index, both associated with global warming \citep{li2011changes}. Additionally, variations in Atlantic SST anomalies, influenced by the AMO, are believed to affect SEUS precipitation through their impact on NASH variability \citep{li2012variation,li2019impacts,pegion2022understanding,zhang2022decadal}.

However, the impact of these interactions on subseasonal SEUS precipitation remains unclear. Most studies at this timescale have focused solely on the variability in the position and intensity of the NASH in modulating SEUS rainfall. The influence of large-scale climate drivers from the Pacific Ocean and their interaction with NASH is still not well understood. Therefore, this study aims to understand the subseasonal variability of SEUS precipitation from both the Atlantic and Pacific Ocean perspectives. We employ a novel application of weather regimes in a conditional approach, called \enquote{Conditional Regimes (CR),} to investigate the mechanisms driving anomalous positive and negative SEUS precipitation. Unlike traditional weather regimes, which identify persistent and recurrent patterns on a given day and their associated impacts on surface temperature and precipitation \citep{robertson1999large,casola2007identifying,amini2019control,nabizadeh2022summertime,lee2023new}, our approach leverages the concept of weather regimes to explore the underlying physical processes more than identifying patterns and their effects. The structure of this paper is as follows: Section 2 outlines the datasets and methodology. Section 3 provides a detailed discussion of the results, and Section 4 concludes with a conclusion and discussion.

\section{Data and methodology}

\subsection{Data and Index calculations}
\begin{figure}[h]
 \centerline{\includegraphics[width=33pc,angle=0]{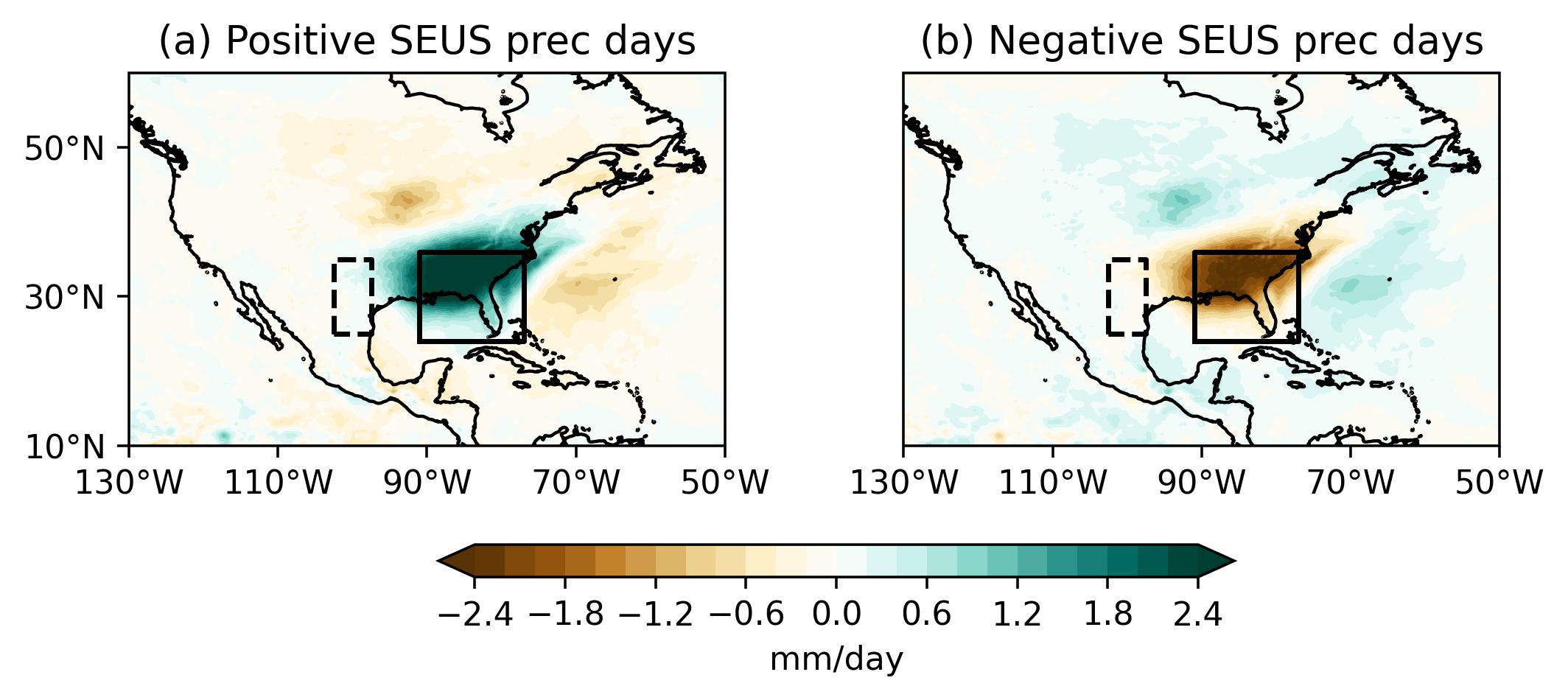}}
  \caption{Spatial composites of precipitation anomalies across North America and surrounding regions, based on positive (\(P' > 0\)) and negative (\(P' < 0\)) rainfall days in the southeastern U.S. (24\degree--36\degree N, 77\degree--91\degree W, indicated by the solid black box) for the period 1979-2019. The dashed box outlines the region (25\degree--35\degree N, 102.5\degree--97.5\degree W) used to calculate the central low-level jet.}\label{f1}
\end{figure}

Our study employs the ERA5 reanalysis dataset from the European Centre for Medium Range Weather Forecasts (ECMWF) \citep{hersbach2020era5}. We analyze daily mean values derived from 6-hourly data (0000, 0600, 1200, and 1800 UTC) with a horizontal resolution of 0.25\degree for the boreal summer season (June-August, JJA) over the period from 1979 to 2019. All atmospheric variables, including geopotential height and winds, as well as SST datasets, are sourced from the ERA5 reanalysis. Moreover, we utilize ERA5-Land data at a spatial resolution of 0.25\degree $\times$ 0.25\degree \citep{munoz2021era5} to calculate daily precipitation anomalies over the SEUS region for the same period.  Daily precipitation anomalies are calculated by subtracting the climatology at each grid point, which is determined as the average value for each day of the year over the entire period. The mean precipitation is then area-averaged over land points within the SEUS region (24\degree--36\degree N, 91\degree--77\degree W, shown in Figure \ref{f1}) and weighted by the cosine of latitude. This calculated precipitation anomaly index is utilized to classify between positive and negative rainfall days. A precipitation anomaly greater than $0$ indicates a positive SEUS precipitation day, while an anomaly less than $0$ denotes a negative SEUS precipitation day. Figure \ref{f1} shows the composite of anomalous positive and negative precipitation days for ERA-5 data to show precipitation patterns across both land and ocean. It highlights the notable variability in the SEUS region during the summer months. The ERA-5 positive and negative precipitation anomaly composites are consistent with the NOAA CPC global unified gauge-based precipitation product \citep{chen2008assessing} for the same time period (Supplement Figure 1). Precipitation in the region can be explained through net moisture convergence using moisture budget theory \citep{chakraborty2021asymmetric, singhai2023indian}. Therefore, we utilize vertically integrated moisture flux divergence from the ERA5 reanalysis to connect the changes in the dynamical circulation with the precipitation.

To investigate the impact of NASH on SEUS rainfall, we compute several indices, such as the central low-level jet (LLJ) and the western ridge of NASH. The Central LLJ index represents the averaged meridional wind at 850 hPa over the region 25\degree--35\degree N, 102.5\degree--97.5\degree W \citep{weaver2011recurrent}, as indicated by the dashed box in Fig. \ref{f1}. The western ridge of NASH is defined as the westernmost point where easterly winds change to westerly winds. Mathematically, this occurs where \( u = 0 \) and \( \frac{\partial u}{\partial y} > 0 \), where \( u \) represents the zonal wind component \citep{liu2004relationship}. Strong convection over the SEUS region is generally governed by the western ridge of the NASH, particularly along the 1560 gpm isoline \citep{li2012variation,wei2019intraseasonal}. Here, we use 1520 gpm instead of the 1560 gpm used by \cite{li2011changes,li2012variation,li2013atmospheric} because we find the maximum difference in the NASH western ridge location for positive and negative precipitation days during the study period using 1520gpm (see Supplementary Fig. 2). We also consider the East Asian precipitation index to measure the impact of the East Asian summer monsoon on SEUS rainfall. This index is defined as the area-averaged anomalous precipitation over Central China and southern Japan (27\degree--35\degree N, 100\degree--140\degree E) \citep{zhu2016new}.

\subsection{Conditional Weather Regimes (CR)}

To differentiate the influences of lower-level and upper-level processes on SEUS precipitation, we compute weather regimes at both levels. This analysis uses daily geopotential height fields at 850 hPa (Z850) and 200 hPa (Z200) obtained from the ERA5 reanalysis \citep{hersbach2020era5} for the June–August period over 41 summers, spanning 1979--2019. We consider Z850 and Z200 to capture variability from both lower and upper atmospheric levels. The selection of a domain for the \enquote{Pacific-North America-Atlantic} (PNAA) region is an important factor contributing to differences among studies of weather regimes over the SEUS. For example, \cite{nabizadeh2022summertime} focuses on the North Pacific and North American regions (0--90\degree N, 150--60\degree W) for clustering analysis, while \cite{molina2023subseasonal} and \cite{robertson2020toward} conducts regime analysis solely over the North American sector (10--70\degree N, 150--40\degree W). In this study, we selected a domain extending from 150\degree E to 20\degree W and 10\degree N to 80\degree N, encompassing nearly all of North America—from the Aleutian Islands to eastern Greenland and from central Mexico to the Canadian Arctic. This domain notably includes portions of both the North Atlantic and North Pacific to capture atmospheric and oceanic variability associated with these oceans. The weather regimes presented here are robust to small changes in the domain size (±10\degree). 

The Z850 and Z200 fields over the PNAA region (150\degree E to 20\degree W, 10\degree N to 80\degree N) are processed by first applying a time filter to remove rapid synoptic-scale transients through running 5-day means. The climatological seasonal cycle is removed by fitting a parabola to the seasonal evolution of 5-day means at each grid point and averaging these parabolas over all years \citep{straus1983role}. The anomaly fields for Z850 and Z200 are then normalized using the square root of the area-averaged temporal standard deviation. Principal component analysis (PCA) is conducted on each vector, with the leading 12 modes accounting for approximately 75\% and 70\% of the total normalized variance, respectively. The principal component (PC) time series serves as coordinates in a 12-dimensional phase space. K-means cluster analysis \citep{diday1976clustering,desbois1982automatic,michelangeli1995weather} is then applied within this PC phase space. This iterative method assigns each state in the phase space (5-day mean) uniquely to one of the \( k \) clusters, where \( k \) is chosen beforehand. Consequently, the anomaly for each 5-day mean is assigned to a specific regime. The value of \( k \) is selected by running the clustering algorithm until convergence for various \( k \) values and identifying the largest \( k \) where the Pearson correlation between the centroid coordinates in PC space is less than zero. This ensures that the maximum number of clusters are all anticorrelated. This threshold is chosen to maximize the number of clusters while minimizing feature overlap among them, aiming to identify the smallest number of clusters necessary to capture the predominant flow patterns. Supplementary Figure 3 illustrates that this approach results in \( k = 4 \) clusters.

\section{\label{results}Results}

\begin{figure}[h]
 \centerline{\includegraphics[width=33pc,angle=0]{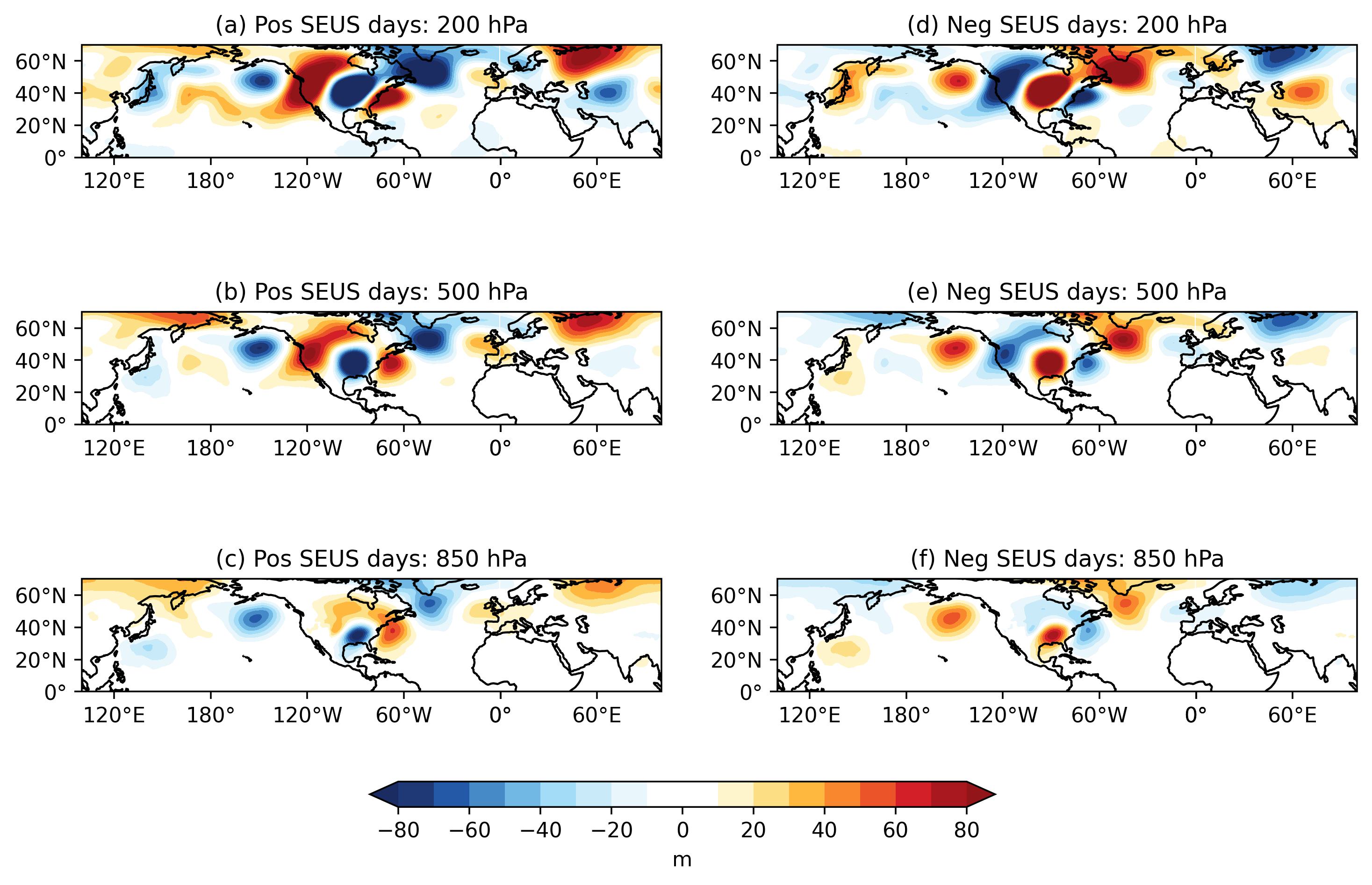}}
  \caption{Mean composite of the anomalous geopotential height at lower (850 hPa), middle (500 hPa), and upper (200 hPa) atmospheric levels for days with positive and negative rainfall events over the SEUS region.}\label{f2}
\end{figure}

To understand the subseasonal rainfall variability in the SEUS region, the primary focus has traditionally been on the displacement of the western ridge of the NASH or SST variability from the Atlantic Ocean. However, the influence of Pacific Ocean variability on SEUS rainfall through various upper-level atmospheric teleconnections is still unexplored and not well understood and documented. Therefore, it is crucial to examine the circulation patterns at both upper and lower levels, involving both the Pacific and Atlantic Oceans, that result in positive and negative rainfall anomalies in the SEUS region. Figure \ref{f2} presents the mean composite of the anomalous geopotential height at lower (850 hPa), middle (500 hPa), and upper (200 hPa) atmospheric levels for days with anomalous positive and negative rainfall.

During positive rainfall days, there is an anomalous decrease in geopotential height at 850 hPa (Fig \ref{f2}c), which steers moist air from the Gulf of Mexico and the Atlantic Ocean into the SEUS. This low-pressure anomaly is further intensified at mid-to-higher levels (500 hPa, shown in Fig. \ref{f2}b, and 200 hPa, shown in Fig. \ref{f2}a), thereby promoting precipitation in the SEUS. Additionally, a wave train pattern originating from the North Pacific Ocean is observed at mid and upper levels, which could further enhance the transport of moisture from the northeastern Pacific Ocean into the SEUS. This configuration of changes in lower-level geopotential height is often associated with the eastward retreat of the western ridge of the NASH \citep{wei2019intraseasonal}, which will be examined in detail in section \ref{pos}\ref{pos_low}. At higher levels, it can be associated with the propagation of the Rossby wave along the subtropical jet stream, reflecting the influence of large-scale climate drivers from the Pacific Ocean, as discussed in the section \ref{pos}\ref{pos_high}.
 
Conversely, during negative rainfall days, the increased geopotential height anomalies at 850 hPa (Fig. \ref{f2}f) over the SEUS are often linked to the more zonally extended and northwestward position of the western ridge of the NASH \citep{li2012variation}. At middle (Fig. \ref{f2}e) and higher (Fig. \ref{f2}d) atmospheric levels, these high-pressure anomalies are enhanced, indicating a diminished moisture influx into the SEUS region. This configuration at both low and higher atmospheric levels suggests subsidence, which inhibits convective activity and results in drier conditions over the SEUS region, as detailed in sections \ref{neg}\ref{neg_low} and \ref{neg}\ref{neg_high}.
 
Our findings suggest that SEUS rainfall dynamics during the summer are influenced not only by low-level convergence driven by the position and strength of the NASH, as shown by previous studies \citep{li2012variation, wei2019intraseasonal}, but also by upper-level dynamics originating from the Pacific Ocean. While earlier research has largely overlooked the significant impact of upper-level factors, this study emphasizes the role of variability arising from both the Atlantic and Pacific Oceans in triggering anomalous precipitation events in the SEUS region. In the next section, we will examine the influence of circulation patterns from both lower and upper atmospheric levels on subseasonal SEUS variability using conditional weather regimes.

\subsection{\label{pos}Anomalous positive SEUS rainfall events:}
\subsubsection{\label{pos_low}Lower-level circulation and anomalous positive rainfall events:}

\begin{figure}[h]
 \centerline{\includegraphics[width=33pc,angle=0]{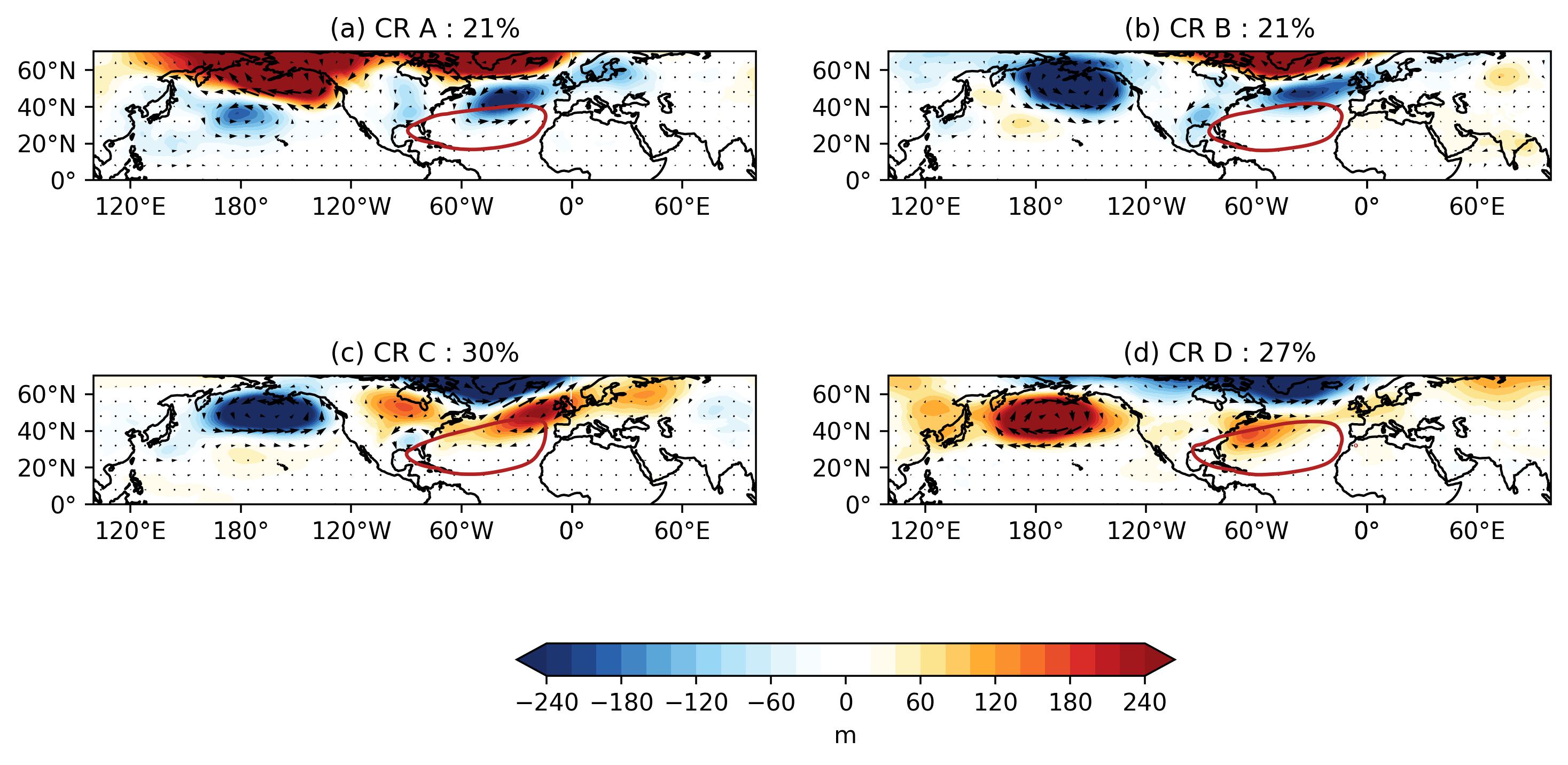}}
  \caption{The averaged anomalous geopotential height at lower atmospheric levels (Z850 hPa) for days with anomalous positive rainfall is shown for each of the four conditional regimes, based on lower-level geopotential height (Z850). The percentage shown above indicates the overall frequency of each regime during anomalous positive rainfall days in the summer from 1979 to 2019. The 1520-gpm isopleth is also shown (red contour) as the NASH boundary. }\label{f3}
\end{figure}

\begin{figure}[h]
 \centerline{\includegraphics[width=33pc,angle=0]{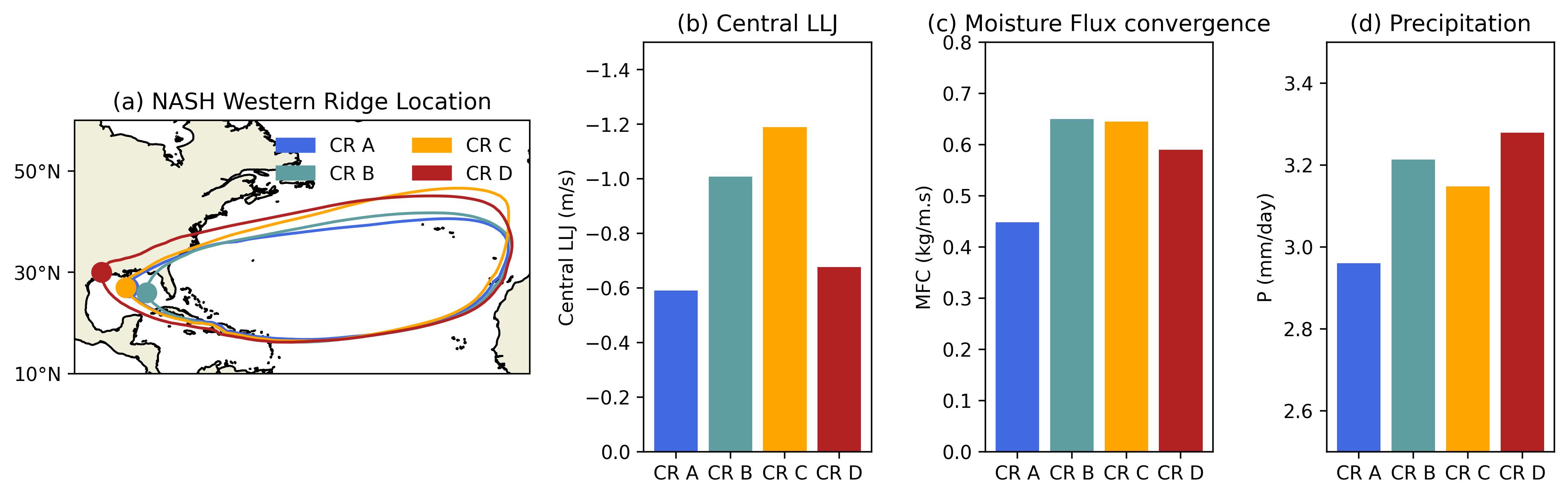}}
  \caption{(a) Composite of the NASH boundary (1520-gpm isopleth, shown by contours) and its western ridge (marked by dots) are depicted for each of the four conditional weather regimes (CR A, CR B, CR C, and CR D), based on lower-level geopotential height (Z850) for anomalous positive rainfall days. The corresponding changes in (b) central low-level jet ($m/s$), (c) moisture flux convergence ($kg/m.s$), and (d) precipitation ($mm/day$) over the SEUS region are also shown.}\label{f4}\end{figure}

The NASH has a significant impact on precipitation patterns in the SEUS by affecting low-level circulation over the Atlantic Ocean \citep{davis1997north,li2013atmospheric}. Here, we perform a conditional weather regime analysis for positive SEUS rainfall days using lower-tropospheric geopotential height anomalies to identify distinct clusters representing persistent and recurring circulation patterns, as illustrated in Figure \ref{f3}, which also delineates the boundary of the NASH using the 1520-gpm isopleth. An important advantage of using weather regimes over composite analysis (as shown in Figure \ref{f2}) is the ability to capture distinct phases of atmospheric circulations that might be averaged out in a standard composite analysis. Each cluster displays distinct phases of the NASH and associated atmospheric circulation patterns that contribute to positive rainfall events in the SEUS region, as described below. This interaction can be understood as a two-way process: the position and intensity of the NASH influence the strength of the central LLJ, leading to increased convergence and rainfall over the SEUS (Figure \ref{f4}). In turn, these rainfall maxima affect the intensity and strength of the NASH.

The first phase is shown by conditional regimes (CR) A and B, which occur 21\% (Fig. \ref{f3}a) and 22\% (Fig. \ref{f3}b) of summer days, respectively, and feature an anomalous trough extending across the upper Midwest and Southeastern US. Furthermore, there is a notable decrease in geopotential height along the northern boundary of the NASH. This decrease is more pronounced in CR B, situated just northeast of the Southeastern US region, compared to CR A. The intensified weakening of NASH in CR B causes its western ridge to retreat eastward, shifting from 89\degree W in CR A to 85\degree W in CR B (see Fig. \ref{f4}a). As a result, the reduced intensity of the NASH reduces the zonal pressure gradient \citep{wei2019intraseasonal}, thereby weakening the central meridional low-level jet (Fig. \ref{f4}b). However, this weakening of the central LLJ is stronger in CR B ($-1 m\ s^{-1}$) compared to CR A ($-0.59 m\ s^{-1}$). The weakened LLJ and the eastward shift of NASH's western ridge induce an anomalous cyclonic circulation over the Southeastern US (Fig. \ref{f3}a and \ref{f3}b). In CR B, this effect is more pronounced than CR A, with a further retreat of the western ridge and a weaker LLJ. This modulation enhances moisture convergence (Fig. \ref{f4}c) across the Southeastern US, leading to an increase in rainfall (Fig. \ref{f4}d) in CR B ($P=3.2\ mm\ day^{-1}$) compared to CR A ($P=2.9\ mm\ day^{-1}$). 

The second phase, represented by Conditional Regimes C and D, is more frequent, occurring on 30\% (Fig. \ref{f3}c) and 27\% (Fig. \ref{f3}d) of summer days, respectively. Both regimes exhibit an anomalous trough centered around the SEUS region, similar to CR A (Fig. \ref{f3}a) and B (Fig. \ref{f3}b). This facilitates the influx of moisture-laden winds from the Gulf of Mexico and the Atlantic Ocean (Fig. \ref{f3}c and \ref{f3}d), promoting convective activity in the SEUS region with rainfall of $3.14 mm/day$ in CR C and $3.27 mm/day$ in CR D, higher than the $2.96 mm/day$ observed in CR A (Fig. \ref{f4}). This increased rainfall in CR C and D leads to excessive condensational heating over the SEUS \citep{wei2019intraseasonal}, triggering a wave train-like pattern located northeast of the SEUS region (Fig. \ref{f3}c and \ref{f3}d), which is absent in CR A and B (Fig. \ref{f3}a and \ref{f3}b). In CR C, the anomalous ridge along the NASH northern boundary is displaced further east than in CR D. This wave train originates from the Southeastern US and propagates eastward over time along the westerly winds to the northern boundary of the NASH. As a result, the impact on the size and intensity of the NASH is less in CR C compared to CR D. In CR D, the western ridge of the NASH expands 4.5\degree westward, shifting its position northwest to 30\degree N, 94.5\degree W, compared to its position in CR C at 27\degree N, 90\degree W (Fig. \ref{f4}a). This shift in the NASH western ridge is consistent with findings by \cite{li2012variation} and \cite{wei2019intraseasonal}, which indicate that enhanced precipitation in the SEUS region modifies the strength and position of the NASH.

This analysis highlights the two-way interaction between lower-level NASH and positive rainfall variability in the SEUS, which may be overlooked in traditional composite analysis (Fig. \ref{f2}). For example, composite analysis misses the weakening along the northern boundary of the NASH (Fig. \ref{f2}c), capturing only one phase of the interaction, whereas a two-way process is at play. Additionally, these features are not fully resolved when weather regimes are applied to all summer or year-round days \citep{nabizadeh2022summertime,lee2023new}. Thus, using conditional weather regime analysis for anomalous positive rainfall days offers a more comprehensive framework for understanding the physical processes driving NASH's role and summer rainfall variability in the SEUS. However, the impact of the NASH is confined to lower levels. To understand the role of upper-level processes in driving SEUS positive rainfall, we conduct a conditional weather regime analysis for upper-level atmospheric variables, which will be discussed in the next section \ref{pos_high}.

\subsubsection{\label{pos_high}Upper-level circulation and anomalous positive rainfall events}

\begin{figure}[h]
 \centerline{\includegraphics[width=35pc,angle=0]{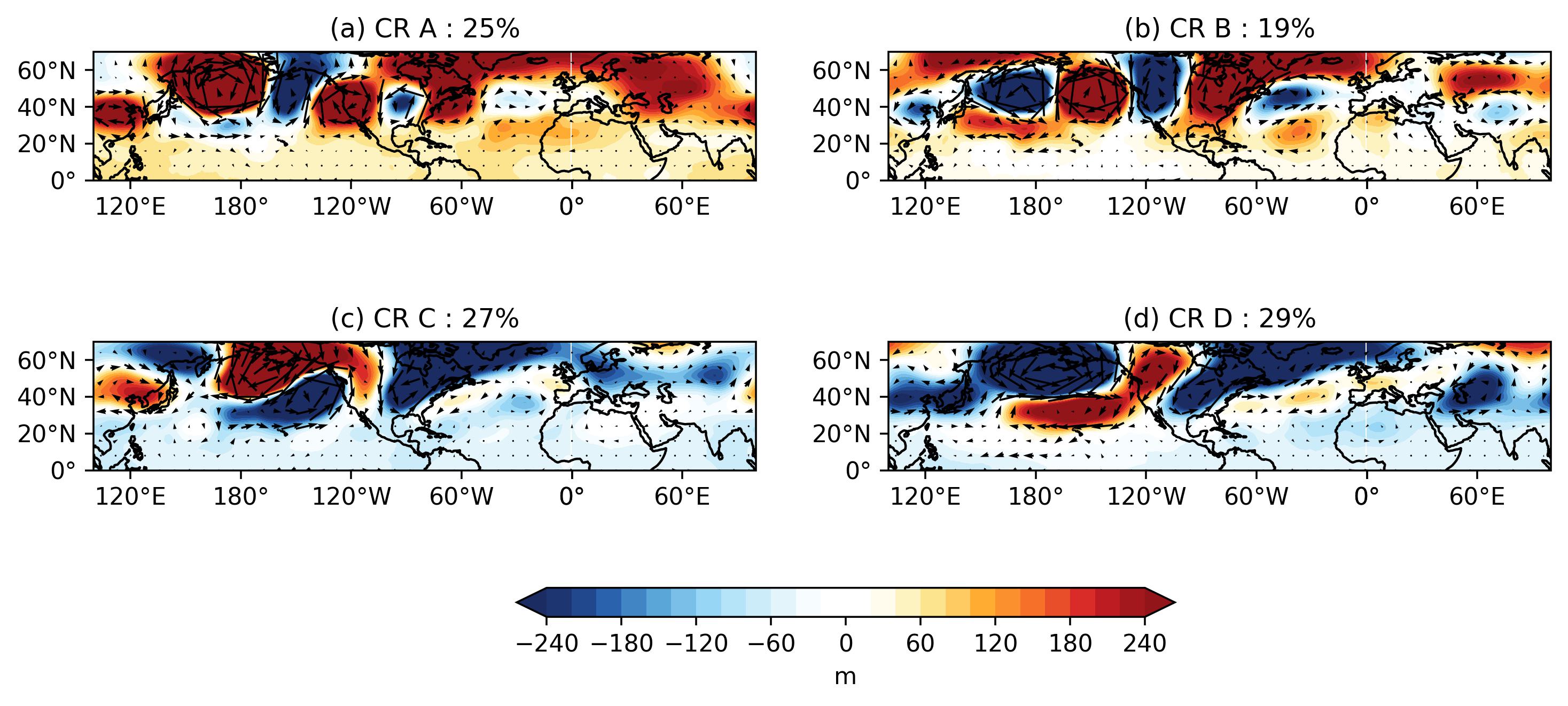}}
  \caption{The averaged anomalous geopotential height at upper atmospheric levels (Z200 hPa) for days with anomalous positive rainfall is shown for each of the four conditional regimes, based on upper-level geopotential height (Z200) for anomalous positive rainfall days. The percentage shown above indicates the overall frequency of each regime during anomalous positive rainfall days in the summer from 1979 to 2019.}\label{f5}
\end{figure}

Variations in SST anomalies in the Pacific and Atlantic oceans can affect SEUS precipitation patterns by altering upper-level circulation. In Figure \ref{f2}a, a small wave train pattern emerging from the North Pacific Ocean is observed, but its source cannot be fully understood through composite analysis. To explore this further, conditional weather regime analysis can be applied to upper atmospheric levels across the Pacific-North America-Atlantic region to determine the role of Pacific and Atlantic climate drivers in driving this wave train, which leads to anomalous positive SEUS rainfall, as shown in Figure \ref{f5}.

The first two regimes, CR A (Fig. \ref{f5}a) and CR B (Fig. \ref{f5}b), which together occur 44\% of the time, are characterized by a planetary wave-train pattern that originates in East Asia and travels to the Continental US. This pattern is not clearly visible in Fig. \ref{f2}a, particularly over the western North Pacific Ocean, due to the superposition of other Pacific Ocean influences. This planetary wave train is referred to as the \enquote{Asian North American teleconnection}, as pointed out by \cite{zhu2016new}. CR A features a Rossby wave train originating from East Asia, characterized by three pairs of anomalous cyclones and anticyclones that link East Asia to SEUS rainfall anomalies. \cite{zhu2016new} performed idealized heating experiments with an atmospheric general circulation model, demonstrating that an increase in rainfall over Central China and southern Japan releases latent heat, leading to the formation of an upper-level anticyclonic anomaly (high-pressure) over northern China and a cyclonic anomaly (low-pressure) east of Japan. This heating initiates a Rossby wave train along the upper-tropospheric westerly jet stream, resulting in an anticyclonic anomaly over the northwest Pacific, a cyclonic anomaly over the northeast Pacific, an anticyclonic anomaly over the northwestern US, a cyclonic anomaly over the central US, and an anticyclonic center over the SEUS region. A similar pattern is also observed in the CR A regime (Fig. \ref{f5}a). In contrast, CR B is characterized by an anomalous trough centered over East Asia, unlike the anomalous ridge seen in CR A. This trough is associated with anomalous diabatic cooling across subtropical East Asia. As a result, the wave train pattern in CR B is the reverse of that in CR A, especially over the Pacific Ocean. CR B exhibits a dipole pattern in geopotential height over the continental United States, which is absent in CR A. This wave train influences rainfall variability in the Southeastern US by affecting the upper-level circulation, which in turn modulates the low-level circulation through thermal wind balance.

\begin{figure}[h]
 \centerline{\includegraphics[width=35pc,angle=0]{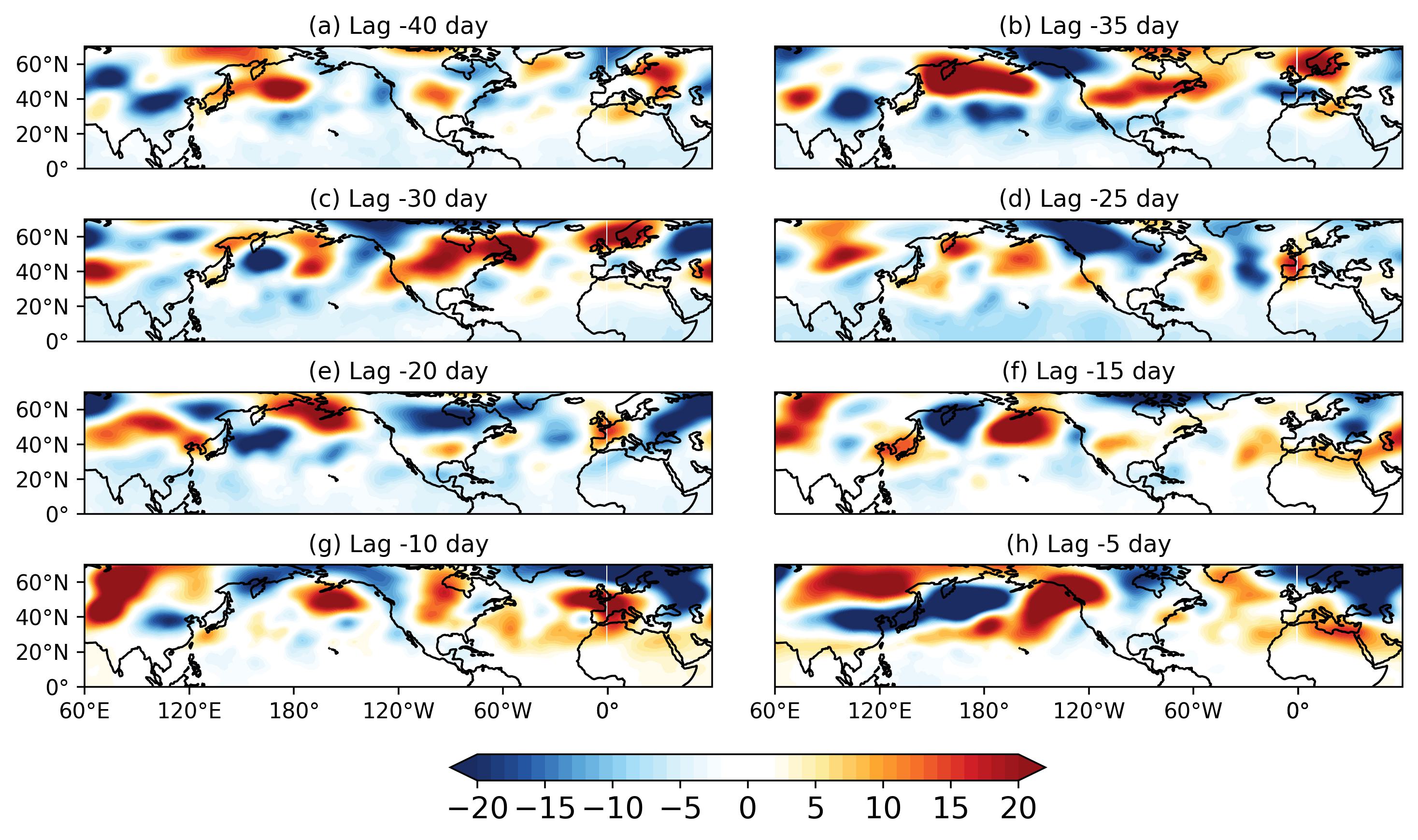}}
  \caption{Time-lag regression patterns as obtained by regressing the 200 hPa geopotential height with the East Asia precipitation index, which is derived by area-averaging precipitation anomalies over Central China and southern Japan (27\degree--35\degree N, 100\degree--140\degree E). The time lags are set to be from -40 day to -5 day (a-h), respectively.}\label{f6}
\end{figure}
 
Additionally, the propagation speed of this wave is influenced by the mean flow of the westerly jet stream, which is slower in summer than in winter. Figure \ref{f6} shows the temporal evolution of this wave train using the time-lag regression of the geopotential height at 200 hPa with the East Asia precipitation index. A similar wave train is noticeable at a 30-day lag (Fig. \ref{f6}c), exhibiting a pattern that closely resembles what is observed in CR B (Fig. \ref{f5}b). This finding aligns with the study by \cite{zhu2016new}, which also identifies that it takes around 30 days (10-40 days) to fully establish the ANA teleconnection pattern and its impact on the continental USA.
 
Another set of conditional regimes, CR C and CR D occur on 27\% (Fig. \ref{f5}c) and 29\% (Fig. \ref{f5}d) of summer days, respectively. CR C is marked by an anomalous trough over the central North Pacific Ocean and a pronounced ridge over the western United States, extending from the North Pacific Ocean. This regime also features a downstream trough over the Southeastern US extending from the North Atlantic Ocean, enhancing the low-level cyclonic circulation and moisture convergence in that region (Supplementary Figure 4c), leading to an increase in precipitation over the SEUS region. In contrast, CR D is characterized by an anomalous ridge over the central North Pacific Ocean and its extension towards the western US, accompanied by a downstream trough over the Southeastern US. This trough is also present in the lower atmospheric levels (Supplementary Figure 4d), forming a barotropic structure over the Southeastern US (SEUS) region. This structure facilitates the flow of warm, moist air from the Gulf of Mexico, enhancing convective activity in the SEUS.

The conditional weather regime of the upper-level geopotential height underscores the impact of Pacific Ocean dynamics on SEUS rainfall. It reveals the planetary Rossby wave train originating from East Asia and extending to the SEUS region, a feature not fully captured in Figure \ref{f2}a. This analysis also shows that this Rossby wave is triggered by anomalous heating or cooling associated with the East Asian monsoon and takes about 30 days to fully develop. Overall, using conditional weather regime analysis for anomalous positive rainfall days offers a comprehensive framework for understanding the physical processes from both the Pacific and Atlantic Oceans driving SEUS rainfall variability.

\subsection{\label{neg} Anomalous negative SEUS rainfall events:}
This section delves into the mechanisms underlying anomalous negative rainfall events in the SEUS region. It utilizes conditional weather regimes based on both lower and upper-level geopotential heights to identify the processes involved. This approach is similar to that used in section \ref{pos}, which investigated positive rainfall events. By examining the interactions and dynamics at different atmospheric levels, this section aims to provide a comprehensive understanding of how these factors contribute to negative rainfall anomalies in the SEUS region.

\subsubsection{\label{neg_low}Lower-level circulation and anomalous negative rainfall events:}

\begin{figure}[h]
 \centerline{\includegraphics[width=33pc,angle=0]{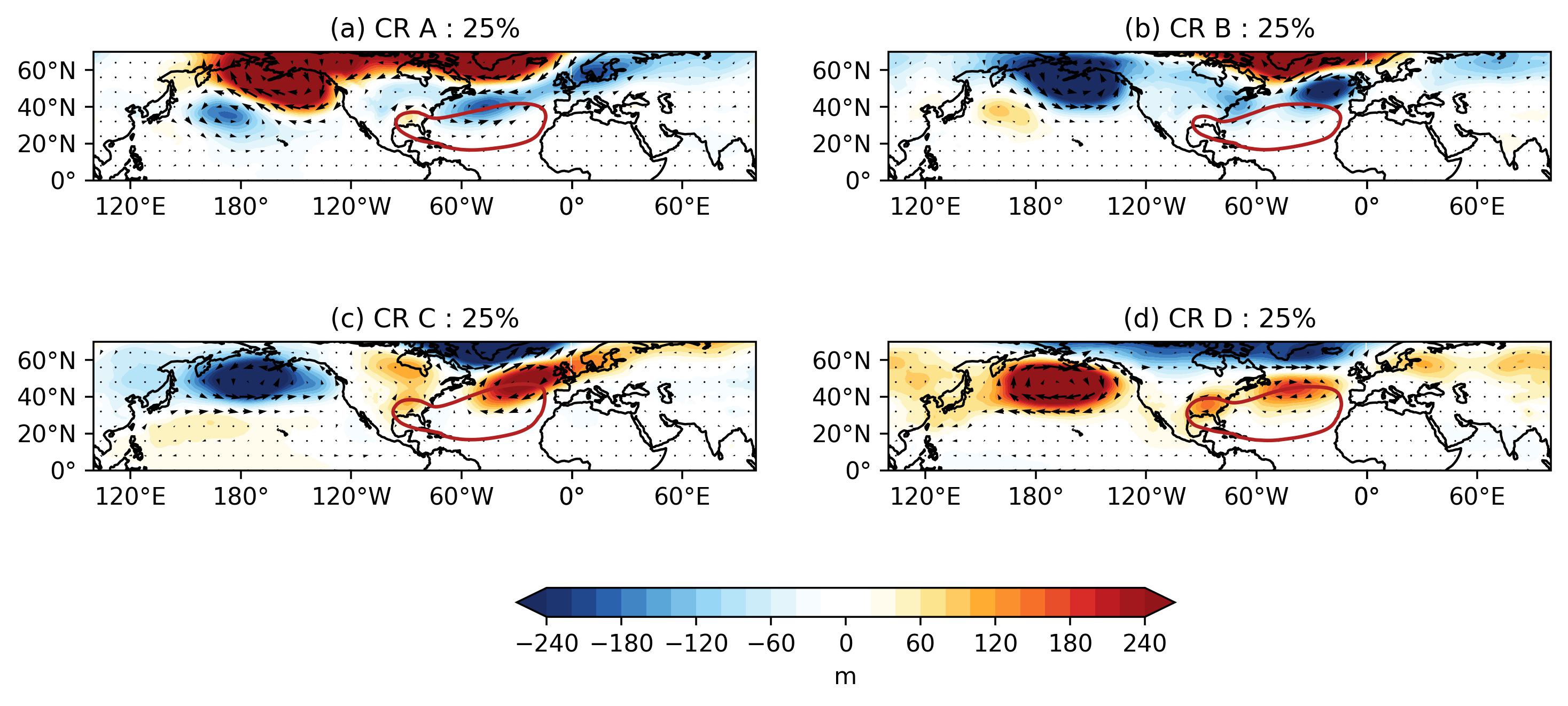}}
  \caption{The averaged anomalous geopotential height at lower atmospheric levels (Z850 hPa) for days with anomalous negative rainfall is shown for each of the four conditional regimes, based on lower-level geopotential height (Z850). The percentage shown above indicates the overall frequency of each regime during anomalous negative rainfall days in the summer from 1979 to 2019. The 1520-gpm isopleth is also shown (red contour) as the NASH boundary. }\label{f7}
\end{figure}

To determine the impact of the NASH's position and strength on anomalous negative rainfall events, Figure \ref{f7} presents a conditional weather regime analysis based on lower tropospheric geopotential height (Z850) for days with anomalous negative rainfall. As noted by \cite{li2012variation}, the northwestward positioning of the NASH relative to its climatological mean, along with its subsequent expansion and the westward shift of its western ridge, are crucial factors in reducing rainfall over the SEUS region. This analysis further investigates the two-way interaction between the NASH's position and intensity and the reduction in rainfall over the SEUS region by identifying persistent and recurring circulation patterns, similar to section \ref{pos_low}.

\begin{figure}[h]
 \centerline{\includegraphics[width=33pc,angle=0]{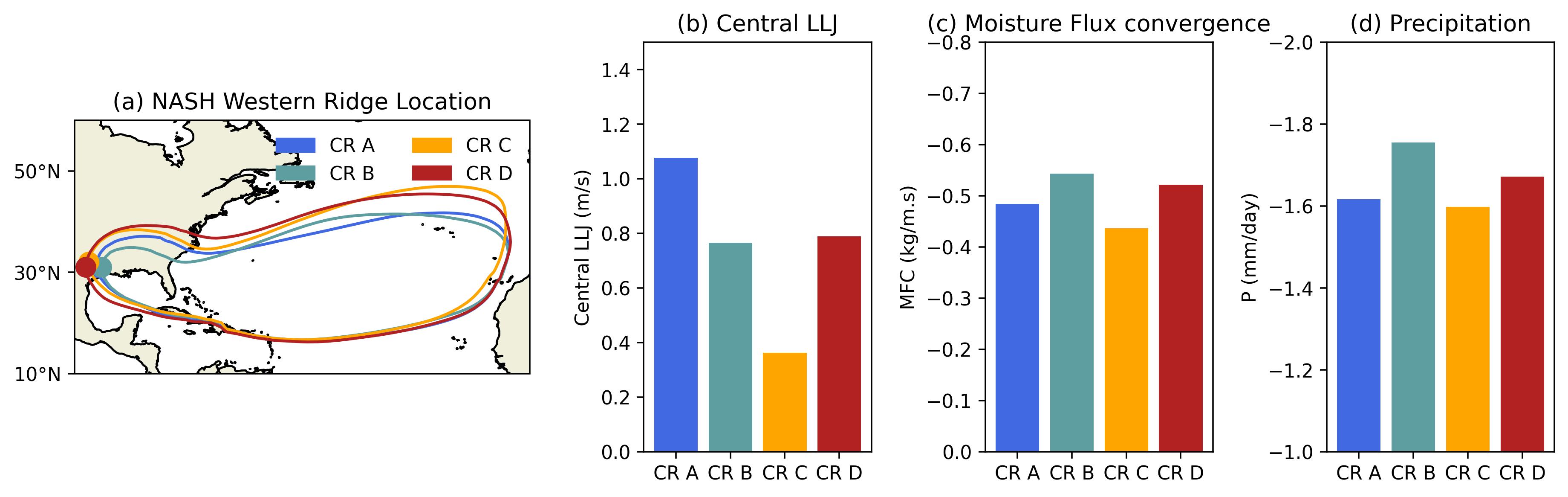}}
  \caption{(a) Composite change in the NASH boundary (1520-gpm isopleth, shown by contours) and its western ridge (marked by dots) are depicted for each of the four conditional weather regimes (CR A, CR B, CR C, and CR D), based on lower-level geopotential height (Z850) for anomalous negative rainfall days. The corresponding changes in (b) central low-level jet ($ m\ s^{-1}$), (c) moisture flux convergence ($kg\ m^{-1}s^{-1}$), and (d) precipitation ($mm\ day^{-1}$) over the SEUS region are also shown.}\label{f8}\end{figure}

Phase 1 is represented by Conditional Regimes CR A (Fig. \ref{f7}a) and CR B (Fig. \ref{f7}b), each accounting for 25\% of days with negative rainfall. Both these regimes are characterized by an anomalous ridge over the SEUS and the North Atlantic Ocean, in contrast to the trough observed during positive rainfall events (Fig. \ref{f3}a and \ref{f3}b). During anomalous negative rainfall events, the NASH western ridge expands, covering much of the SEUS region (Fig. \ref{f8}a). Although the western ridge of NASH is similarly positioned in both regimes—at 32\degree N, 97\degree W in CR A and 31\degree N, 97.5\degree W in CR B (Fig. \ref{f8}a)—the influence of NASH is less pronounced in CR A. This reduced impact in CR A is due to the presence of the anomalous trough along the northern boundary of NASH (Fig. \ref{f7}a). Consequently, the increase in the zonal pressure gradient \citep{wei2019intraseasonal} leads to the strengthening of the central low-level jet in CR A ($LLJ=1.0 \ m\ s^{-1}$) and CR B ($LLJ=0.79 \ m\ s^{-1}$) (Fig. \ref{f8}b). The enhanced LLJ and the westward shift of NASH's western ridge induce an anomalous anticyclonic circulation over the Southeastern US (Fig. \ref{f7}a and \ref{f7}b). In CR A, this effect is less pronounced compared to CR B due to the weakening of the NASH along its northern boundary. This modulation restricts the transport of moist air from the Gulf of Mexico and the Atlantic into the SEUS, leading to reduced convective activity (Fig. \ref{f8}c) and lower precipitation in CR B ($P=-1.7 \ mm\ day^{-1}$) compared to CR A ($P=-1.6 \ mm\ day^{-1}$) (Fig.~\ref{f8}d).

Another phase shown by Conditional regimes CR C (Fig. \ref{f7}c) and CR D (Fig. \ref{f7}d) occurs on 25\% and 25\% of summer days with anomalous negative rainfall, respectively. Both regimes exhibit an anomalous ridge centered around the SEUS region, similar to regimes CR A (Fig. \ref{f7}a) and CR B (Fig. \ref{f7}b). This ridge hinders the influx of moisture-laden winds from the Gulf of Mexico and the Atlantic Ocean, reducing convective activity (Fig. \ref{f8}c) in the SEUS region with a rainfall decrease of $1.6\ mm\ day^{-1}$ in CR C and $1.65\ mm\ day^{-1}$ in CR D (Fig. \ref{f7}d). The decreased rainfall in regimes CR C and CR D leads to anomalous cooling over the SEUS \citep{wei2019intraseasonal}, leading to the formation of a wave train-like pattern located northeast of the SEUS region (Fig. \ref{f7}c and \ref{f7}d), which is absent in regimes CR A and CR B. This wave train originates from the Southeastern US and propagates eastward along the northern boundary of the NASH. In CR C, this anomalous ridge is displaced further east than in CR D. Consequently, the impact on the size and intensity of the NASH is more significant in CR D compared to CR C. In CR D, the size of the western ridge of the NASH expands by 1\degree westward, shifting its position southeast to 31\degree N, 94.5\degree W, compared to its position in CR C at 31\degree N, 93.5\degree W. This northwest tilt of the NASH western ridge is consistent with findings by \cite{li2012variation}, which indicate that such a position reduces precipitation in the SEUS region.

Similar to positive rainfall events, this analysis uncovers the two-way interaction between lower-level NASH and negative rainfall variability in the SEUS. By capturing both phases of this relationship, conditional weather regime analysis for anomalous negative rainfall days offers a clearer understanding of how the NASH drives dry conditions in the region.

\subsubsection{\label{neg_high}Upper-level circulation and anomalous negative rainfall events:}

\begin{figure}[h]
 \centerline{\includegraphics[width=35pc,angle=0]{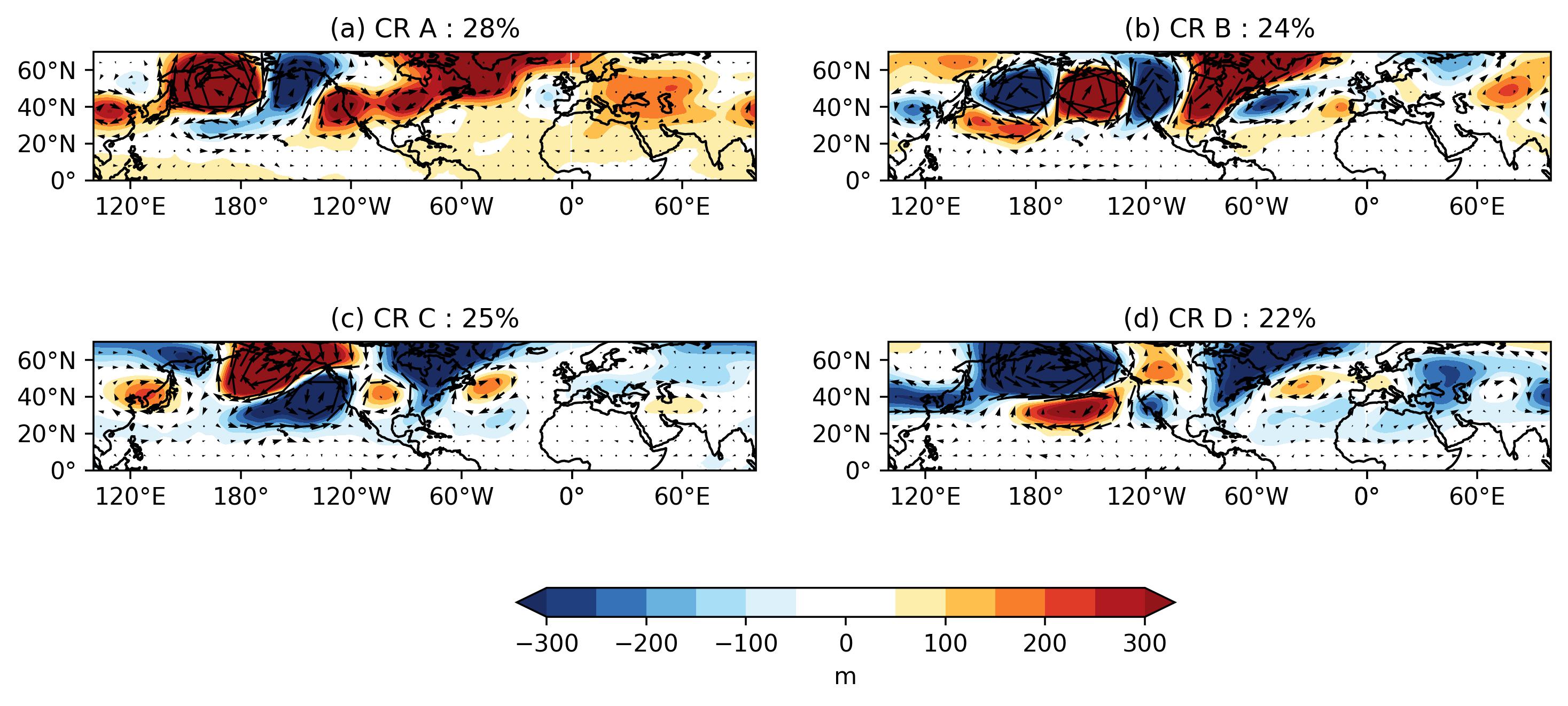}}
  \caption{The averaged anomalous geopotential height at upper atmospheric levels (Z200 hPa) for days with anomalous negative rainfall is shown for each of the four conditional regimes, based on upper-level geopotential height (Z200). The percentage shown above indicates the overall frequency of each regime during anomalous negative rainfall days in the summer from 1979 to 2019.}\label{f9}
\end{figure}

Figure \ref{f9} shows a conditional weather regime analysis of upper-tropospheric geopotential height for anomalous negative rainfall days, similar to Figure \ref{f5}. This analysis highlights the role of upper-level processes in driving dry conditions in the SEUS region.

Conditional regimes CR A (Fig. \ref{f9}a) and CR B (Fig. \ref{f9}b), occurring 28\% and 24\% of the time, are characterized by a pronounced planetary wave-train pattern that originates from East Asia and propagates towards North America (similar to Fig. \ref{f5}a and \ref{f5}b). These Rossby waves are triggered by anomalous heating or cooling associated with East Asian subtropical monsoon rainfall, forming part of the ANA teleconnection pattern, as described by \cite{zhu2016new}.  CR A has an anomalous ridge centered over East Asia, while CR B  features an anomalous trough in the same region. Both regimes display upper-level patterns with alternating high and low-pressure systems, showing notable similarity over the Pacific Ocean but with opposite signs. Over the continental US, the patterns diverge significantly. In CR A, this results in a coherent rise in geopotential height over the continental US due to the influence of the North Atlantic Ocean, whereas in CR B, the ANA wave train originates from East Asia and extends to the southeastern US. This leads to high-pressure anomalies over the SEUS, causing subsidence and dry conditions in both CR A and CR B.

Conditional regimes CR C (Fig. \ref{f9}c) and CR D (Fig. \ref{f9}d) occur 25\% and 25\% of the time. CR C displays a strong cyclonic anomaly over the central North Pacific, extending into the western US. This pattern leads to an anomalous ridge over the central US, accompanied by a downstream trough over the SEUS. This trough in the SEUS region is mainly associated with the influence of the North Atlantic Ocean. It promotes subsidence, reducing cloud cover and precipitation in the SEUS region. CR D is marked by a strengthening over the central North Pacific Subtropical High region. This CR D also features an extensive ridge extending from the North Pacific into the northwestern US and Canada. A downstream trough over the western US and SEUS region leads to significant subsidence over the SEUS, inhibiting convective activity and leading to reduced rainfall.

In summary, anomalous negative rainfall events over the SEUS region are primarily influenced by large-scale atmospheric teleconnections from the upper levels and NASH from the lower levels. One key factor is the ANA wave train, which originates from East Asia and propagates toward North America along the westerly jet stream. This wave train is often triggered by anomalous heating or cooling associated with East Asian subtropical monsoon rainfall, leading to the development of alternating high and low-pressure systems. Additionally, the expansion of the NASH from the lower levels plays a significant role in modulating moisture transport from the Gulf of Mexico and the Atlantic Ocean. The NASH's influence on low-level circulation patterns contributes to subsidence and the formation of high-pressure anomalies over the SEUS, suppressing precipitation. Together, these teleconnections create atmospheric conditions that result in reduced rainfall and drier days in the SEUS region.

\section{Conclusions}

This study explores the mechanisms behind anomalous positive and negative rainfall events in the SEUS region during summer using ERA5 reanalysis data from 1979 to 2019. It focuses on the interplay between large-scale atmospheric teleconnections and the NASH. Unlike previous research, which primarily focuses on the position and intensity of the NASH, this study provides a broader perspective by considering additional climatic drivers. To achieve this, we employ a conditional weather regime analysis using upper and lower tropospheric geopotential height to account for processes and phenomena from both levels that influence SEUS precipitation variability. We perform weather regime analysis over the Pacific-North America-Atlantic region to incorporate variability from both the Pacific and Atlantic oceans. Our analysis identifies four distinct clusters of persistent and recurring circulation patterns that might be overlooked in traditional composite analyses. The findings underscore the significant roles that upper-level teleconnections and lower-level NASH modulation play in governing precipitation across this region.

Our analysis reveals how different phases of the NASH's position and strength influence subseasonal rainfall in the SEUS. Using conditional weather regimes of lower-level geopotential height during positive and negative rainfall events, we identify four regimes representing two distinct phases of the NASH and SEUS rainfall. One phase shows when the northern boundary of the NASH weakens and shifts eastward, diminishing the central low-level jet and inducing stronger cyclonic circulation over the SEUS, leading to enhanced moisture convergence and higher SEUS rainfall. In another phase, this increase in rainfall triggers a wave train pattern originating from the SEUS, traveling along the northern boundary of the NASH and subsequently strengthening the NASH. Conversely, on negative rainfall days, when the NASH strengthens and shifts westward, it reinforces the central low-level jet and induces anticyclonic circulation, resulting in decreased moisture convergence and reduced rainfall. Another phase features an anomalous ridge over the SEUS, leading to reduced rainfall due to diminished moisture influx and consequently weakening the NASH.

The conditional weather regime analysis of upper-level geopotential height highlights the substantial influence of Pacific Ocean dynamics on precipitation variability in the SEUS. This analysis reveals that variability in SEUS precipitation from the Pacific is mainly linked to Rossby waves, which are generated by anomalous heating or cooling associated with East Asian subtropical monsoon rainfall. These Rossby waves travel along the westerly jet stream from East Asia to North America, and their propagation significantly impacts rainfall patterns across the SEUS. Additional variability in SEUS rainfall can be attributed to the dynamics of the North Atlantic Oceans, leading to anomalous trough or ridge over the SEUS region. This, in turn, alters the advection of moist air from the Gulf of Mexico and the Atlantic Ocean, resulting in an increase or decrease in precipitation. 

In summary, the conditional weather regime analysis highlights the intricate interplay between oceanic and atmospheric processes from the Pacific and Atlantic Oceans, revealing the complex mechanisms that drive summertime subseasonal precipitation variability in the SEUS. The combined influence of the NASH's position at lower levels, along with upper-level effects from the North Pacific Ocean, contributes significantly to the variability in SEUS rainfall. This underscores the importance of gaining a thorough understanding of these factors to enhance subseasonal predictions in this region.

%

%

\clearpage
\acknowledgments
The authors thank ECMWF for providing ERA5 daily atmospheric data sets. K.P. acknowledges funding from the NOAA/WPO grant (NA210AR4320204). 


%
%
\datastatement

The ERA5 data used in the study is obtained from the ECMWF (https://www.ecmwf.int/en/forecasts/datasets/browse-reanalysis-datasets). 


%






%



\bibliographystyle{ametsocV6}
\bibliography{references}

\end{document}